\documentclass[preprint,12pt]{elsarticle}
\usepackage{multirow}
\usepackage{setspace}
\usepackage{multirow}
\usepackage{latexsym, amsbsy, amssymb}
\usepackage{epsfig}
\usepackage{amsmath,amsthm,color,amsbsy}

\def\R{{\mathbb R}}
\newcommand{\indep}{\;\, \rule[0em]{.03em}{.67em} \hspace{-.25em}
\rule[0em]{.65em}{.03em} \hspace{-.25em}
\rule[0em]{.03em}{.67em}\;\,}

\newcommand{\X}{{\mathbf X}}
\newcommand{\Y}{{\mathbf Y}}
\newcommand{\Z}{{\mathbf Z}}
\newcommand{\T}{{\mathbf T}}

\newcommand{\B}{{\mathbf B}}
\newcommand{\A}{{\mathbf A}}
\newcommand{\C}{{\mathbf C}}
\newcommand{\D}{{\mathbf D}}

\newcommand{\K}{{\mathbf K}}

\newcommand{\tX}{{\tilde \X}}
\newcommand{\tY}{{\tilde \Y}}
\newcommand{\tZ}{{\tilde \Z}}

\newcommand{\spc}{{\mathcal S}_{\Y|\X}}
\newcommand{\spco}{{\mathcal S}_{Y|\X}}
\newcommand{\spcz}{{\mathcal S}_{Y|\Z}}
\newcommand{\spci}{{\mathcal S}_{Y_i|\X}}
\newcommand{\spcyz}{{\mathcal S}_{\Y|\Z}}

\def\S{{\mathbb S}}

\newcommand{\bb}{\mathbf{b}}
\newcommand{\be}{\boldsymbol{\epsilon}}
\newcommand{\bmu}{\boldsymbol{\mu}}

\newcommand{\bu}{\mathbf{u}}

\newcommand{\E}{\mathrm{E}}
\newcommand{\Var}{\mathrm{Var}}
\newcommand{\Sig}{\boldsymbol{\Sigma}}
\newcommand{\Sigx}{\boldsymbol{\Sigma}_\X}

\newcommand{\bLambda}{\boldsymbol{\Lambda}}
\newcommand{\Sigxy}{\boldsymbol{\Sigma}_{\X\Y}}
\newcommand{\bGamma}{\boldsymbol{\Gamma}}

\newcommand{\bomega}{\boldsymbol{\omega}}

\newcommand{\0}{\bf{0}}

\newcommand{\bxi}{\boldsymbol{\xi}}

\newcommand{\M}{{\mathbf M}}

\newcommand{\spn}{\mathrm{span}}
\newcommand{\rank}{\mathrm{rank}}

\newcommand{\pms}{{\tiny \mathrm{PMS}}}

\newcommand{\pr}{{\tiny \mathrm{PR}}}

\newcommand{\nn}{{\tiny \mathrm{NN}}}

\newcommand{\CR}{{\tiny \mathrm{CR}}}

\newcommand{\wire}{{\tiny \mathrm{WIRE}}}

\newcommand{\kmir}{{\tiny \mathrm{KMIR}}}

\newcommand{\ols}{{\tiny \mathrm{OLS}}}

\newcommand{\rr}{{\tiny \mathrm{RR}}}

\newcommand{\env}{{\tiny \mathrm{ENV}}}

\newcommand{\pls}{{\tiny \mathrm{PLS}}}

\newcommand{\V}{{\mathbf V}}
\newcommand{\bv}{{\mathbf v}}
\newcommand{\bG}{{\mathbf G}}

\newcommand{\I}{{\mathbf I}}

\newcommand{\y}{{\mathbf y}}

\newtheorem{proposition}{{\bf Proposition}}

\newcommand{\bc}{{\mathbf c}}
\newcommand{\bd}{{\mathbf d}}

\newcommand{\ba}{{\mathbf a}}
\newcommand{\bm}{{\mathbf m}}

\newcommand{\bg}{{\mathbf g}}

\newcommand{\W}{{\mathbf W}}

\newcommand{\bff}{{\mathbf f}}

 \journal{""}
 
\begin{document}
\begin{frontmatter}
\title{A selective review of sufficient dimension reduction for multivariate response regression}
\author[label1]{Yuexiao Dong\corref{cor1}}
\address[label1]{Department of Statistics, Operations, and Data Science, Temple University,
Philadelphia, PA, US, 19122}
 \ead{ydong@temple.edu}
 \cortext[cor1]{Corresponding author.}
 \author[label2]{Abdul-Nasah Soale}
\address[label2]{Department of Applied and Computational Mathematics and Statistics, University of Notre Dame, South Bend,
 IN, US, 46556}
 \author[label1]{Michael D. Power}


\begin{abstract}
We review sufficient dimension reduction (SDR) estimators with multivariate response in this paper. A wide range of SDR methods are characterized as inverse regression SDR estimators or forward regression SDR estimators. The inverse regression family include pooled marginal estimators, projective resampling estimators, and distance-based estimators.  Ordinary least squares, partial least squares, and semiparametric SDR estimators, on the other hand, are discussed as estimators from the forward regression family.

\end{abstract}
\end{frontmatter}

\section{Introduction}

For $q$-dimensional response $\Y$ and
$p$-dimensional predictor $\X$, sufficient dimension reduction (SDR) aims to find $\B\in\R^{p\times
d}$ with the smallest possible column space such that
\begin{align}
\label{sdr}
\Y\indep \X \mid \B^\top \X,
\end{align}
 where $\indep$ means independence.  The column space of $\B$, or $\spn(\B)$, is known as the central space, and is denoted as $\spc$.
  The dimension of the central space is referred to as the structural dimension. Denote the columns of $\B$ as $\bb_j$, $j=1,\ldots,d$.
  For a continuous response $\Y$, a regression form of (\ref{sdr}) is
\begin{align}
\label{sdr2}
 \Y=\bg(\bb_1^\top \X,\ldots,\bb_d^\top \X,\be),
 \end{align}
 where $\be$ is $r$-dimensional random error independent of $\X$ (with $r\ge 1$), and 
  $\bg:\R^{d+r}\mapsto \R^q$ is an unknown link function.
 Given i.i.d. samples $\{(\Y_i,\X_i):i=1,\ldots,n\}$ generated from model (\ref{sdr2}), multivariate response SDR focuses on estimating the indices $\bb_1,\ldots,\bb_d$ without necessarily estimating the link function $\bg$.
 
 Since the seminal works of Li (1991) and Cook (1998), many SDR methods have been proposed in the literature. Most of these methods focus on the univariate response case with $q=1$. Existing SDR review papers and a recent SDR book follow a similar trend,  discussing almost exclusively methods  for the univariate response. See, for example,  Yin (2010), Ma and Zhu (2013), Li (2018), and Dong (2021). This paper aims to fill in this gap and provides a selective review of SDR with multivariate response.  
 The rest of the paper is organized as follows. In section 2, we review multivariate response SDR methods through inverse regression. Forward regression methods for multivariate response SDR are discussed in section 3.  We conclude the paper with some emerging trends in section 4. 
Without loss of generality, we assume  $\E(\X)=\0$ and $\E(\Y)=\0$ throughout the paper.

\section{Multivariate response SDR through inverse regression} 
 
\subsection{SIR and slicing-based inverse regression methods} 
Denote $\Var(\X)=\Sigx$ and the standardized predictor as $\Z=\Sigx^{-1/2}\X$. First we review the original SIR with univariate response $Y\in \R$. Under the following linear conditional mean (LCM) assumption
\begin{align}
\label{lcm0}
\E(\X\mid\B^\top \X) \mbox{ is linear in }\B^\top \X, \mbox{ where }\B \mbox{ is the basis of }\spco,
 \end{align}
 Li (1991) showed that $\E(\Z\mid Y)\in \spcz$.   Due to an equivariant property of the central space in Theorem 2.2 of Li (2018),  we have 
 \begin{align}
\label{sir1}
\Sigx^{-1}\E(\X\mid Y)=\Sigx^{-1/2}\E(\Z\mid Y)\in \spco.
\end{align}
 Denote $\bxi_h^{(0)}=\Sigx^{-1}\E(\X\mid Y\in J_h)$ for $h=1,\ldots,H$, where $J_1,\ldots,J_H$ is a partition of the support of $Y$.
 Let 
  \begin{align*}
\M^{(0)}=\sum_{h=1}^H p_h^{(0)} \bxi_h^{(0)} {\bxi_h^{(0)}}^\top, \mbox{ where }p_h^{(0)}=\E(Y\in J_h).
\end{align*} 
We refer to $\M^{(0)}$ as the SIR kernel matrix. 
In the case of categorical response, the categories become a natural partition. For continuous response,
quantile slicing is used for the partition in the original SIR. In particular, for $\ell=1,\ldots, H-1$, let $\tau_\ell$ be  
 the $\ell H^{-1}$-th population quantile of $f_Y$, the density function of $Y$. The partition used in SIR becomes $(-\infty,\tau_1)$, $(\tau_1,\tau_2)$, $\ldots$, $(\tau_{H-1},\infty)$. Instead of the continuous response $Y$, we now have the discretized response $\tilde Y=\sum_{h=1}^H I(Y\in J_h)$, where $I(\cdot)$ denotes the indicator function. 
 According to Theorem 2.3 of Li (2018), it can be shown that 
  \begin{align}
\label{sir2}{{\mathcal S}_{\tilde Y|\X}}\subseteq \spco.
\end{align}
Applying (\ref{sir1}) to $\tilde Y$, we have 
   \begin{align}
\label{sir3}\bxi_h^{(0)}\in {{\mathcal S}_{\tilde Y|\X}}.
\end{align}
  Equations (\ref{sir2}) and  (\ref{sir3}) lead to $\spn(\M^{(0)})\subseteq \spco$. 
 Denote $\hat{\M}^{(0)}$ as the sample version of $\M^{(0)}$. SIR then uses the eigenvectors corresponding to the $d$ leading eignevalues of 
$\hat\M^{(0)}$ to recover the central space $\spco$.

Many slicing-based inverse regression methods  have been proposed in the literature after the introduction of the original SIR. We discuss the extensions that are relevant for multivariate response methods to be reviewed in later sections. To synthesize the intraslice means across different slices, the original SIR takes an eigenvalue decomposition approach and can be suboptimal in terms of asymptotic efficiency. Cook and Ni (2005) suggested a minimum discrepancy approach and proposed the inverse regression estimator with optimal  asymptotic efficiency. 
Note that $\M^{(0)}$ is directly related to $\bLambda_{\mathrm{I}}=\Var(\E(\Z \mid \tilde Y))$ through  $\M^{(0)}=\Sigx^{-1/2}\bLambda_{\mathrm{I}} \Sigx^{-1/2}$. Let 
$$\bLambda_{\mathrm{II}}=\E\left\{[\Var(\Z\mid \tilde Y)-\E(\Var(\Z\mid \tilde Y))]^2\right\}$$
Under LCM and  an additional constant conditional variance (CCV) assumption that $\Var(\X\mid\B^\top \X) $ is nonrandom, Li (1991) showed that 
$\spn(\bLambda_{\mathrm{II}})\subseteq \spcz$, and this becomes known as the SIR-II method. For $\alpha\in[0,1]$, SIR-$\alpha$ (Li, 1991) uses $(1-\alpha)\bLambda_{\mathrm{I}}^2+\alpha \bLambda_{\mathrm{II}}$ 
to recover the central space. While the original SIR only uses information from the intraslice means, SIR-II and SIR-$\alpha$ synthesizes both the intraslice means and the intraslice variances. 
Other slicing-based inverse regression methods include sliced average variance estimation (SAVE) (Cook and Weisberg, 1991), covariance inverse regression estimator (CIRE) (Cook and Ni, 2006), and directional regression (Li and Wang, 2007).

 \subsection{Pooled marginal estimators for multivariate response SDR} 

In the original SIR, we use quantile slicing to partition the support of the univariate response. Direct analogy of this strategy no longer works in the case of multivariate response $\Y=(Y_1,\ldots,Y_q)^\top$. For example, if each marginal response is dichotomized through median split, the support of $\Y\in\R^q$ is then partitioned into $2^q$ hypercubes.  As $q$ increases, the number of observations within each hypercube decreases
exponentially, which deteriorates the estimation efficacy. Pooled marginal estimators do not have this limitation, as we demonstrate in this section.

Note that 
$\Y\indep \X \mid \B^\top \X$ in (\ref{sdr}) implies $Y_i\indep \X \mid \B^\top \X$, $i=1,\ldots,q$. From the definition of central space,
 we  have the following observation.
\begin{proposition}\label{prop1}
For $i=1, \ldots,q$, $\spci\subseteq\spc$.
\end{proposition}
\noindent Proposition \ref{prop1} is the key to pooled marginal slicing estimators and pooled marginal estimators in general. It implies that we can combine univariate response SDR estimators of $\spci$ to get the multivariate response SDR estimator.

Before we state the pooled marginal slicing method,
we need the modified LCM assumption
\begin{align}
\label{lcm}
\E(\X\mid\B^\top \X) \mbox{ is linear in }\B^\top \X, \mbox{ where }\B \mbox{ is the basis of }\spc.
 \end{align} 
 Compared to (\ref{lcm0}), note that (\ref{lcm}) accounts for multivariate response. Unless specified otherwise, LCM refers to (\ref{lcm}) hereafter. We also state the modified CCV assumption as
 \begin{align}
\label{ccv}
\Var(\X\mid\B^\top \X) \mbox{ is nonrandom}, \mbox{ where }\B \mbox{ is the basis of }\spc.
 \end{align} 
 For $i=1, \ldots,q$, let $J_{i,1},\ldots,J_{i,H}$ be a partition of the support of $Y_i$. Denote $$\bxi_{i,h}=\Sigx^{-1}\E(\X\mid Y_i\in J_{i,h}), p_{i,h}=\E(Y_i\in J_{i,h}) \mbox{, and } \M_i=\sum_{h=1}^H p_{i,h} \bxi_{i,h} \bxi_{i,h}^\top.$$
Under LCM,  we have $\spn(\M_i)\subseteq\spci\subseteq\spc$.  The pooled marginal slicing (PMS) (Aragon, 1997) defines the weighted sum of $\M_i$ as
   \begin{align*}
 \M_\pms=\sum_{i=1}^q w_i \M_i. 
  \end{align*} 
Then we have $\spn(\M_\pms)\subseteq\spc$.  The eigenvectors corresponding to the $d$ leading  eignevalues of $\hat\M_\pms$, the sample version of $\M_\pms$, can be used to recover $\spc$. Aragon (1997) suggested to choose $w_i$ either as equal weights or proportional to the leading eigenvalues of $\M_i$. Pooled marginal slicing avoids partitioning of the overall support of the multivariate response and the ``curse of dimensionality'', as only marginal response slicing is involved.   Lue (2009) studied the asymptotic properties of the sample PMS estimator, and extended the sequential test approach in Li (1991) to decide the structural dimension of $\spc$.  Note that $\M_\pms$ is based on combining $q$ original SIR kernel matrices $\M_i\in\R^{p\times p}$. Pooled marginal slicing that combines $q$ kernel matrices of SIR-$\alpha$ is studied in Saracco (2005) and Barreda  et al. (2007). SAVE with pooled marginal slicing is discussed in Yoo et al. (2010). Coudret et al. (2014) proposed a new marginal slicing method that combines the leading eigenvectors of $\M_i$ instead of $\M_i$ itself.

For univariate response SDR methods that do not require slicing such as principal Hessian directions (PHD) (Li, 1992) and central K-th moment space estimation (CKMS) (Yin and Cook, 2002), Proposition 
\ref{prop1} can be applied directly to get the pooled marginal estimators of $\spc$. Extensions of CKMS to multivariate response SDR are studied in Cook and Setodj (2003) and Yin and Bura (2006), and PHD with multivariate response is discussed in Lue (2010). 
Following the minimum discrepancy approach in Cook and Ni (2005), Yoo and Cook (2007)  combined marginal CKMS estimators optimally to achieve asymptotic efficiency. 

Similar to their univariate response counterparts, multivariate response SDR methods that are related to SIR and SIR-$\alpha$ require the LCM assumption (\ref{lcm}), while multivariate response methods that are based on SAVE and PHD  require both the LCM assumption (\ref{lcm}) and the CCV assumption (\ref{ccv}).

\subsection{Projective resampling for multivariate response SDR}

Consider the following two statements: (i)
$\Y\indep \X \mid \B^\top \X$; (ii) for any fixed vector $\bv\in\R^q$,
 $\bv^\top\Y\indep \X \mid \B^\top \X$. It can be shown that the two statements imply each other. This observation is the basis of the projective resampling (PR) method proposed in Li et al. (2008). The next result is adapted from Theorem 3.1 of Li et al. (2008). 
 
 \begin{proposition}\label{prop2}
Suppose $\V$ is a random vector uniformly distributed on unit sphere $\S^q$. For each realization $\bv\in\R^q$, let $\M(\bv)\in\R^{p\times p}$ be a positive semi-definite matrix such that $\spn[\M(\bv)]={{\mathcal S}_{\bv^\top\Y|\X}}$. Then  $\spn\{\E[\M(\V)]\}=\spc$.
\end{proposition}
\noindent In Proposition \ref{prop1}, the conclusion implies that the pooled marginal estimators are unbiased and may only recover  a proper subset of the full central space.  Proposition \ref{prop2}, on the other hand, states that the projective resampling estimators can exhaustively recover the central space $\spc$ as long as the estimation of  ${{\mathcal S}_{\bv^\top\Y|\X}}$ is exhaustive for any fixed $\bv$.

In practice, we take an i.i.d. sample $\bv_{(1)},\ldots,\bv_{(m_n)}$ from a uniform distribution on $\S^p$. For example, we can take $\bv_{(j)}=\bG_{(j)}/\|\bG_{(j)}\|$. Here $\|\cdot\|$ denotes the Euclidean norm and  $\bG_{(1)},\ldots,\bG_{(m_n)}$ are i.i.d. $N(\0,\I_p)$. Let $J_{(j),1},\ldots,J_{(j),H}$ be a partition of the support of $\bv_{(j)}^\top\Y$.
For $j=1,\ldots,m_n$, denote 
\begin{align*}
&\bxi_{(j),h}=\Sigx^{-1}\E(\X\mid \bv_{(j)}^\top\Y\in J_{(j),h}), p_{(j),h}=\E(\bv_{(j)}^\top\Y\in J_{(j),h}), \mbox{ and }\\
&\hspace{1in}\M_{(j)}=\sum_{h=1}^H p_{(j),h} \bxi_{(j),h} \bxi_{(j),h}^\top.
\end{align*}
The projective resampling SIR kernel matrix becomes
$$\M_\pr=\frac{1}{m_n}\sum_{j=1}^{m_n}\M_{(j)}$$
At the sample level, the estimator of $\M_\pr$  is denoted as $\hat\M_\pr$.  PR-SIR then uses eigenvectors corresponding to the $d$ leading eigenvalues of $\hat\M_\pr$ to recover the central space. Under some regularity assumptions, Theorem 3.2 of Li et al. (2002) states that $\hat\M_\pr$ is a $\sqrt{n}$ consistent estimator of $\E[\M(\V)]$ as long as $n=O(m_n)$. In addition to SIR, 
projective resampling can be combined with other univariate SDR methods.  Li et al. (2002) discussed PR-SAVE. Distance covariance  (DCOV) (Sz{\'e}kely et al., 2007) for SDR with univariate response is studied in Sheng and Yin (2013, 2016), and PR-DCOV is proposed for multivariate response SDR in Chen et al. (2019).

\subsection{Distance-based methods for multivariate response SDR}
Contour regression (Li et al., 2005) is originally proposed for univariate response SDR, and it can be easily adapted for multivariate response. Let $(\tX,
\tY)$ be an independent copy  of $(\X,\Y)$. Denote
$\Z=\Sigx^{-1/2}\X$, $\tZ=\Sigx^{-1/2}\tX$, and
$\A(\varepsilon)=\E[(\Z-\tZ)(\Z-\tZ)^\top\mid \|\Y-\tY\|<\varepsilon]$. 
Let $\bLambda_\CR=(2\I_p-\A(\varepsilon))^2$.
 Under LCM (\ref{lcm}) and CCV (\ref{ccv}), it can be shown that $\spn(\bLambda_\CR)\subseteq \spcyz$. See, for example, Theorem 6.1 in Li (2018). Denote $\M_\CR=\Sigx^{-1/2}\bLambda_\CR\Sigx^{-1/2}$. The equivariant property of the central space leads to $\spn(\M_\CR)\subseteq \spc$.

At the sample level, let $\{(\Y_i,\X_i):i=1,\ldots,n\}$ be an i.i.d. sample. Denote $\hat{\Sig}_\X$ as the sample variance of $\X$, $\hat{\bmu}_\X$ as the sample mean of $\X$, and $\hat\Z_i={\hat{\Sig}_\X}^{-1/2}(\X_i-\hat{\bmu}_\X)$. For a given $\varepsilon>0$, compute the index set   \begin{align}\label{dis1}
G(\varepsilon)=\{(i,j):1\le i<j\le n,\|\Y_i-\Y_j\|<\varepsilon\}.
\end{align}
Let $|G(\varepsilon)|$ be the cardinality of $G(\varepsilon)$.
Then $\A(\varepsilon)$ can be estimated by 
 $$\hat \A(\varepsilon)=|G(\varepsilon)|^{-1}\sum_{(i,j)\in G(\varepsilon)}(\hat\Z_i-\hat\Z_j)(\hat\Z_i-\hat\Z_j)^\top,$$
 and $\M_\CR$ is estimated by 
 $$\hat \M_\CR={\hat{\Sig}_\X}^{-1/2}(2\I_p-\hat\A(\varepsilon))^2 {\hat{\Sig}_\X}^{-1/2}.$$
The eigenvectors corresponding to the $d$ leading eigenvalues of $\hat\M_\CR$ are the final contour regression estimator.

An important step in the sample version estimation of contour regression is to get index set $G(\varepsilon)$ in (\ref{dis1}), which directly depends on the pairwise Euclidean distance between two responses $\Y_i$ and $\Y_j$. Nearest neighbor inverse regression (NNIR) (Hsing, 1999) is another method that uses distances between responses. In particular, for a fixed index $i$, let $i^*$ be the response index that corresponds to the nearest neighbor of $\Y_i$ such that 
$$\|\Y_{i^*}-\Y_i\|=\min_{1\le j\le n, j\neq i} \|\Y_j -\Y_i\|.$$
The sample level kernel matrix for NNIR is then defined as
$$\hat\M_\nn={\hat{\Sig}_\X}^{-1/2} \hat \bLambda_\nn {\hat{\Sig}_\X}^{-1/2}, \mbox{ where }\hat \bLambda_\nn =\frac{1}{2n}\sum_{i=1}^n(\hat \Z_i\hat \Z_{i^*}^\top+\hat \Z_{i^*}\hat \Z_{i}^\top).$$
The corresponding population level kernel matrix is
$$\M_\nn=\Sigx^{-1/2} \bLambda_\nn \Sigx^{-1/2}, \mbox{ where } \bLambda_\nn= \Var(\E(\Z\mid\Y)).$$
From Theorem 3 and Lemma 5 of Hsing (1999), we know that $\hat\M_\nn$ is a $\sqrt{n}$ consistent estimator of $\M_\nn$. Under the LCM assumption (\ref{lcm}), it is easy to see that $\spn(\M_\nn)\subseteq \spc$.

Ying and Yu (2020) proposed Fr{\'e}chet SDR with metric-spaced valued response, and their weighted inverse regression ensemble (WIRE) estimator can be directly adapted for multivariate response SDR. In particular, the kernel matrix for the WIRE is
$$\M_\wire=\Sigx^{-1/2} \bLambda_\wire \Sigx^{-1/2}, \mbox{ where } \bLambda_\wire=-\E(\Z\tZ^\top \|\Y-\tY\|).$$
Note that $\bLambda_\wire$ can be reexpressed as
$$\bLambda_\wire=-\E (\E(\Z\mid \Y)\E^\top(\tZ\mid\tY)\|\Y-\tY\|), $$
which is a distance-weighted average of $\E(\Z\mid \Y)\E^\top(\tZ\mid\tY)$. Under the LCM assumption (\ref{lcm}),  we can easily show that $\spn(\M_\wire)\subseteq \spc$.

\subsection{Other inverse regression methods for multivariate response SDR}

Following
  similar arguments for equation (\ref{sir1}), we have $\Sigx^{-1}\E(\X\mid \Y) \in\spc$ under the LCM assumption (\ref{lcm}). Instead of slicing the response to estimate $\E(\X\mid \Y)$, Bura and Cook (2001) imposed a parametric model between response $\X$ and predictor $\bff(\Y)$, where $\bff:\R^q\mapsto \R^m$ is a known function, and 
model $\E(\X\mid \Y)$ as a linear function of $\bff(\Y)$.

In classical SIR, quantile slicing is used to partition the support of the response. K-means inverse regression (KMIR)  (Setodji and Cook, 2004) extends this idea and sets the sample level observations into $K$ groups through K-means clustering (Hartigan, 1975) of the $n$ responses. In particular, denote $C_k$ as the $k$th response cluster and $n_k$ as the cluster size  for $k=1,\ldots, K$. Then $\bar\Z_k=n_k^{-1}\sum_{\Y_i\in C_k}\hat \Z_i$ becomes the intra-cluster mean of the standardized predictor, and the kernel matrix of sample level KMIR is
$$\hat\M_\kmir={\hat{\Sig}_\X}^{-1/2}\hat \bLambda_\kmir {\hat{\Sig}_\X}^{-1/2}, \mbox{ where }\hat \bLambda_\kmir =\frac{1}{n}\sum_{k=1}^K n_k \bar\Z_k \bar\Z_k^\top.$$ 
KMIR is based on SIR, and can be easily extended to other slicing-based inverse regression methods. For example,  
Wen et al. (2009) combined CIRE with K-means clustering, and
SAVE with K-means clustering is discussed in Yoo et al. (2010). 

Denote $m(\y)=\E(\Z\mid \Y=\y)$ and let $f_\Y(\y)$ be the density function of $\Y$. For $\bomega\in\R^q$, the Fourier transformation of $m(\y)f_\Y(\y)$ becomes
$$\psi(\bomega)=\int e^{\iota \bomega^\top \y}m(\y)f_\Y(\y) d\y=a(\bomega)+b(\bomega)i.$$
Here $\iota^2=-1$ is the imaginary unit, $a(\bomega)$ and $b(\bomega)$ are the real part and the imaginary part of $\psi(\bomega)$, respectively. Under LCM, it can be shown that $\spn\{a(\bomega),b(\bomega)\}\subseteq \spcyz$. It turns out that $\psi(\bomega)$ can be simplified as $\E(e^{\iota \bomega^\top \Y}\Z)$, which can be estimated easily at the sample level. Fourier transformation   for univariate response SDR is first discussed in Zhu and Zeng (2006), and  its multivariate response extensions  include Zhu et al. (2010), Weng and Yin (2018), and Wang et al. (2021).

\section{Multivariate response SDR through forward regression}

\subsection{Multivariate response regression under link violation}

Li and Duan (1989) made an interesting discovery about univariate response regression under link violation. In particular, consider a special case of model (\ref{sdr2}) as
$ Y=\bg(\bb^\top\X,\epsilon)$, where $Y$ is univariate response, $\bg(\cdot)$ is an unknown bivariate link function, and $\epsilon$ is independent of $\X$. Li and Duan (1989) suggested that we can assume the unknown link function is linear, and proceed with ordinary least squares estimation to get 
$\bb_\ols=\Sigx^{-1}\E(\X Y)$. Then under the LCM assumption that $\E(\X\mid \bb^\top\X)$ is linear in $\bb^\top\X$, we have $\bb_\ols=c \bb$ for some $c\in\R$. Using the terminology of SDR, this is to say $\bb_\ols\in\spco$. 

For $\Y\in\R^q$, $\X\in\R^p$, $\B\in\R^{p\times q}$, and $\be\in\R^q$ independent of $\X$, the classical multivariate response linear regression model is
\begin{align}\label{mlr}
\Y=\B^\top\X+\be.
\end{align}
The corresponding least squares estimation aims to minimize $\E(\be^\top\be)$ over $\B$, which leads to
$\B_\ols=\Sigx^{-1}\Sigxy$ with $\Sigxy=\E(\X\Y^\top)$. Under model (10), it is easy to see that $\B_\ols=\B$. On the other hand, suppose the true model is (\ref{sdr}) or (\ref{sdr2}). Following similar arguments in Li and Duan (1989) or Theorem 8.3 in Li (2018), we still have $\spn(\B_\ols)\subseteq\spc$ 
under the LCM assumption (\ref{lcm}). 

\subsection{Reduced rank regression, envelopes, and partial least squares}

Let $\A\in\R^{q\times m}$, $\D\in\R^{m\times p}$, and  $m<\min(p,q)$. The reduced rank regression model (Izenman, 1975) considers
\begin{align}\label{rrr}
\Y=\B^\top\X+\be, \mbox{ where }\B^\top=\A\D \mbox{ and }\rank(\A)=\rank(\D)=m.
\end{align}
An important difference between (\ref{mlr}) and (\ref{rrr}) is that $\B$ has full rank in (\ref{mlr}) and reduced rank 
$\rank(\B)=m$ in  (\ref{rrr}).  For a positive definite matrix $\bGamma\in\R^{q\times q}$, we may minimize $\E(\be^\top \bGamma\be)$ over
$\A\in\R^{q\times m}$ and $\D\in\R^{m\times p}$ under the constraint that $\rank(\A)=\rank(\D)=m$. From Theorem 1 of Izenman (1975), we know the solution of this minimization problem leads to 
$$\B_\rr=\Sigx^{-1}\Sigxy\bGamma^{1/2} \left(\sum _{j=1}^m \bu_j\bu_j^\top\right)\bGamma^{-1/2}, $$
where $\bu_j$ is the eigenvector corresponding to the $j$th leading eigenvalue of $\bGamma^{1/2} \Sigxy^\top \Sigx^{-1} \Sigxy\bGamma^{1/2}$.
If we choose $\bGamma=\{\Var(\Y)\}^{-1}$, Theorem 2 of Izenman (1975) states that $\B_\rr$ is directly related to the canonical correlation analysis (CCA) (Hotelling, 1936). CCA for univariate response SDR and  model-free variable selection are studied in Zhou and He (2008) and Alothman et al. (2018), respectively. 
It is easy to see that $$\spn(\B_\rr)\subseteq\spn(\Sigx^{-1}\Sigxy)=\spn(\B_\ols).$$
If we replace model (\ref{rrr}) with model (\ref{sdr}), then we still have $\spn(\B_\rr)\subseteq\spc$ under the LCM assumption.

Denote ${\mathcal S}$ as a subspace of $\R^p$ and let  ${\mathcal S}^\bot$ be its orthogonal complement in $\R^p$. Cook et al. (2013) considered the envelope regression model
\begin{align}\label{env}
\Y=\B^\top\X+\be, \mbox{ where }\spn(\B)\subseteq{\mathcal S}, \Sigx{\mathcal S} \subseteq{\mathcal S} \mbox{ and }\Sigx{\mathcal S}^\bot \subseteq{\mathcal S}^\bot.
\end{align}
Model (\ref{env}) is known as the predictor envelope model, which is an important extension of the original response envelope regression model in Cook et al. (2010). For an excellent review of response envelope regression and its various extensions, please refer to Cook (2018).  Model (\ref{env}) implies that $\mathcal S$ is a reducing subspace of $\Sigx$ that contains $\spn(\B)$. The intersection of all such reducing subspaces is known as the $\Sigx$-envelope of $\spn(\B)$, and is denoted by 
${\mathcal E}_{\Sigx}(\B)$.

Under model (\ref{env}), we have $\B_\ols=\Sigx^{-1}\Sigxy=\B$. Together with Proposition 2.4 of  Cook et al. (2010), we have 
$${\mathcal E}_{\Sigx}(\B)={\mathcal E}_{\Sigx}(\B_\ols)={\mathcal E}_{\Sigx}(\Sigxy).$$
Let $\C\in\R^{p\times m}$ be a semi-orthogonal basis of ${\mathcal E}_{\Sigx}(\Sigxy)$, and we define 
\begin{align}\label{env1}
\B_\env=\C(\C^\top \Sigx \C)^{-1}\C^\top \Sigxy.
\end{align}
Following similar argument from Proposition 5 of Cook et al. (2013), we have $\B_\env=\B$ under model (\ref{env}).  Furthermore, we still have $\B_\env=\B_\ols$ if  we replace model (\ref{env}) with model (\ref{sdr}). It follows that $\spn(\B_\env)\subseteq\spc$ under the LCM assumption.

Envelope model (\ref{env}) is closely related to partial least squares (PLS) (Helland, 1988). For integer $a$, define the multivariate Krylov matrix as
 $$\K_a=\left(\Sigxy,\Sigx\Sigxy,\Sigx^2\Sigxy,\ldots,\Sigx^{a-1}\Sigxy \right)\in\R^{p\times aq},$$
 and denote ${\mathcal K}_a=\spn(\K_a)$.
Cook et al. (2013) showed that there exists an integer $b$ such that ${\mathcal K}_a$ is strictly increasing in $a$ until $a=b$, ${\mathcal K}_a$ becomes a constant for all $a\ge b$, and the constant is ${\mathcal E}_{\Sigx}(\Sigxy)$, the $\Sigx$-envelope of $\spn(\Sigxy)$. 
Similar to  (\ref{env1}), we define
\begin{align}\label{pls}
\B_\pls=\K_a(\K_a^\top \Sigx \K_a)^{-1}\K_a^\top \Sigxy.
\end{align}
It follows that $\B_\pls=\B_\env$ for large enough $a$, and  $\spn(\B_\pls)\subseteq \spc$ under LCM. Note that the PLS estimator in (\ref{pls}) is exactly parallel to the univariate response PLS estimation for dimension reduction in Naik and Tsai (2000). The popular SIMPLS algorithm for PLS (de Jong, 1993) is closely related to $\B_\pls$, and is demonstrated to be still applicable without assuming linear link functions between $\Y$ and $\X$ (Cook and Forzani, 2021).

\subsection{Multiresponse SDR through  semiparametric  regression}
In the case of univariate response, Xia et al. (2002) considered 
a nonparamatric model $Y=m(\B^\top \X)+\epsilon$ for some $\B\in\R^{p\times d}$ and unknown link function $m:\R^{d}\mapsto \R$.
Clearly we have $\spc=\spn(\B)$ under this model.
 At the sample level, for $i=1,\ldots, n$, $m(\B^\top \X_i)$ is approximated by
$$\hat m(\B^\top \X_i)=m(\B^\top \X_j)+\dot{m}^\top(\B^\top \X_j)\B^\top (\X_i-\X_j)=a_j+\bd_j^\top \B^\top(\X_i-\X_j),$$
where $\dot{m}(\bc)={\partial m(\bc)}/{\partial \bc}\in\R^d$, 
$a_j=m(\B^\top \X_j)$, and $\bd_j=\dot{m}(\B^\top \X_j)$. Then we define objective function 
\begin{align}\label{mave}
\sum_{j=1}^n\sum_{i=1}^n \{Y_i-a_j-\bd_j^\top \B^\top(\X_i-\X_j)\}^2 w_{ij},
\end{align}
where $ w_{ij}=K_h(\X_i-\X_j)/\sum_{\ell=1}^n K_h(\X_\ell-\X_j)$ satisfies $\sum_{i=1}^n w_{ij}=1$ and $K_h$ denotes a kernel function with bandwidth $h$. MAVE then proceeds to estimate $\B$ by minimizing the objective function in (\ref{mave}) over $\{a_j,\bd_j\}_{j=1}^n$ and $\B$ under the constraint $\B^\top \B=\I_d$.

For multivariate response, Yin and Li (2011) studied $f_\T(\Y)$ as a family of transformations such that $\T$ is a random vector and $f_\T(\Y)=f_\T(\Y,1)+\iota f_\T(\Y,1)$. To fix the idea, we may take $f_\T(\Y)=e^{\iota \T^\top \Y}$ as the characteristic function. Let $\T_1,\ldots,\T_r$ be i.i.d. copies of $\T$. The ensemble MAVE then minimizes 
\begin{align*}
\sum_{\ell=1}^2\sum_{k=1}^r\sum_{j=1}^n\sum_{i=1}^n \{f_{\T_k}(\Y_i,\ell)-a_{jk}(\ell)-\bd_{jk}^\top(\ell) \B^\top(\X_i-\X_j)\}^2 w_{ij}\rho_j
\end{align*}
to recover the central space. Here $\rho_j$ is a trimming function that excludes some unreliable observations, and we omit its detailed form here. 

For $\Y\in\R^q$ and $\B\in\R^{p\times d}$, Zhang (2021) considered $\Y=\bm(\B^\top \X)+\be$, where $\bm(\cdot)=(m_1(\cdot),\ldots,m_q(\cdot))^\top$. Let $\ba_j=\bm(\B^\top \X_j)\in\R^q$ and $\D_j=(\dot{m}_1(\B^\top \X_j),\ldots,\dot{m}_q(\B^\top \X_j))\in\R^{d\times q}$. Then $\bm(\B^\top \X_i)$ can be approximated by 
$\ba_j+\D_j^\top\B^\top(\X_i-\X_j)$, and multiresponse MAVE proceeds to minimize 
\begin{align*}
\sum_{j=1}^n\sum_{i=1}^n \{\Y_i-\ba_j-\D_j^\top \B^\top(\X_i-\X_j)\}^\top\W \{\Y_i-\ba_j-\D_j^\top \B^\top(\X_i-\X_j)\}w_{ij}\rho_j.
\end{align*}
A natural choice of the weight matrix $\W\in\R^{q \times q}$  is the inverse of $\E(\be\be^\top)$, which can be estimated by its sample counterpart.  Zhu and Zhong (2015) proposed a similar approach, where $\bm(\B^\top \X_i)$ is estimated by  leave-one-out kernel regression and $\B$ is reparameterized such that its first $d$ rows form an identity matrix. An efficient semiparametric estimator under this model is provided in Zhang et al. (2017), and we omit the details here. The advantage of the semiparametric estimators  in this section is that they no longer require the LCM or the CCV assumption, but these estimators are computationally more expensive than the inverse regression estimators and the forward regression estimators that bypass the estimation of the unknown link function $\bm(\cdot)$.

\section{Conclusions}

The dimension reduction methods reviewed in this paper date back to as early as Hotelling (1936), and yet they remain relevant in modern multivariate analysis. 
For example, in the recent Jubilee volume celebrating the $50$th anniversary of {\it Journal of Multivariate Analysis}, two articles are directly related to SDR. 
Girard et al. (2022) reviewed extensions of sliced inverse regression, and one such extension is pooled marginal sliced inverse regression. Among many multivariate methods, Cook (2022) discussed the conceptual connections between SDR, partial least squares, and envelopes. 
In the presence of increasingly complex data, the SDR assumption of dependence of multivariate
response variable with respect to only a few linear combinations of the predictors can help data visualization and facilitate  data analysis. In this paper,  SDR methods with multivariate response are summarized in the inverse regression family and the forward regression family. 

There are some emerging trends in the SDR literature with regards to multivariate response regression.  Ghosh (2022) cast SDR under the information-theoretic framework, and argued that the central space can be viewed as an information bottleneck.  An existing multivariate response SDR method that falls into this framework is Xue et al. (2018). In applications such as missing data analysis, causal inference, and graphical models,  a natural assumption is that response variables interact with each other only through the predictors. Luo (2022)  further assumed that the interactions between the response variables only depend on a few linear combinations of the predictors. We expect to see further development of multivariate response SDR along these directions.


\begin{thebibliography}{99} 

\bibitem{} 
Alothman, A., Dong, Y. and Artemiou, A. (2018). On dual model-free variable selection with two groups of variables. {\it Journal of Multivariate Analysis}, {\bf 167}, 366--377.

\bibitem{} 
Aragon Y. (1997). A Gauss implementation of multivariate sliced inverse regression. {\it Computational  Statistics}, {\bf 12},
355--372.

		\bibitem{} 
			Barreda, L., Gannoun, A. and Saracco, J. (2007).
Some extensions of multivariate sliced inverse regression. 
{\it Journal of Statistical Computation and Simulation}, {\bf 77}, 1--17.	
 
 
 		\bibitem{} 
Bura, E. and Cook, R. D. (2001). Estimating the structural dimension of regressions via parametric inverse regression. {\it Journal of the Royal Statistical 
Society, Series B}, {\bf 63}, 393--410. 
 
 \bibitem{} Chen, X., Yuan, Q. and Yin, X. (2019). Sufficient dimension reduction via distance covariance with
multivariate responses. {\it Journal of Nonparametric Statistics}, {\bf 31}, 268--288.  
 
 \bibitem{} Cook, R. D. (1998). {\it Regression Graphics: Ideas for
Studying Regressions through Graphics}. New York: Wiley.
 
 \bibitem{} Cook, R. D. (2018). {\it An Introduction to Envelopes: Dimension Reduction for Efficient Estimation in Multivariate Statistics}. New Jersey: Wiley.  
 
 \bibitem{} Cook, R.D. (2022).  A slice of multivariate dimension reduction. {\it Journal of Multivariate Analysis}, {\bf 188}, 104812. 
 
  \bibitem{}
 Cook, R.D. and Forzani, L. (2021). PLS regression algorithms in the presence of nonlinearity. {\it Chemometrics and Intelligent Laboratory Systems}, {\bf 213}, 104307.
 
   \bibitem{}
Cook, R. D., Helland, I. S. and Su, Z. (2013).
Envelopes and partial least squares regression. {\it Journal of the Royal Statistical Society, Series B},  {\bf 75}, 851--877. 
 
 
 
  \bibitem{}
 Cook, R. D., Li, B. and Chiaromonte, F. (2010). Envelope models for parsimonious and efficient
 multivariate regression (with discussion). {\it Statistica Sinica}, {\bf 20}, 927--1010. 
 
\bibitem{} Cook, R. D. and Ni, L. (2005).
Sufficient dimension reduction via inverse regression: a minimum discrepancy approach. {\it Journal of the American
Statistical Association}, {\bf 100}, 410--428. 
 
 \bibitem{} Cook, R. D. and Ni, L. (2006).
Using intraslice covariances for improved estimation of the
central subspace in regression. {\it Biometrika}, {\bf 93}, 65--74.
 
\bibitem{}
Cook, R. D. and Setodji, C. M. (2003). A model-free test for reduced rank in multivariate regression. \textit{Journal of the American Statistical Association}, {\bf 98}, 340--351.

\bibitem{}
Cook, R. D. and  Weisberg, S. (1991).  Comment on ``Sliced inverse regression for dimension reduction''. \textit{ Journal of American Statistical Association}, {\bf 86}, 28--33. 
 
 \bibitem{}
de Jong, S. (1993). SIMPLS: an alternative approach to partial least squares regression. {\it Chemometrics and Intelligent Laboratory Systems}, {\bf 18}, 251--263. 
 
 
\bibitem{}
Coudret, R., Girard, S. and Saracco, J. (2014). A new sliced inverse regression method for multivariate response. Computational Statistics and Data Analysis, {\bf 77}, 285--299.  
 
\bibitem{} Dong, Y. (2021). A brief review of linear sufficient dimension reduction through optimization. {\it Journal of Statistical Planning and Inference}, {\bf 211}, 154--161.   
 
  \bibitem{}
Ghosh, D. (2022). Sufficient dimension reduction: an information-theoretic viewpoint. {\it Entropy}, {\bf 24}, 167. 
 
 \bibitem{}
Girard, S., Lorenzo, H. and Saracco, J. (2022). Advanced topics in sliced inverse regression. {\it Journal of Multivariate Analysis}, {\bf 188}, 104852. 
 
  \bibitem{}
Hartigan, J. A. (1975). {\it Clustering Algorithms}. New York: Wiley. 
 
   \bibitem{}
Helland, I. S. (1988). On the structure of partial least squares regression. {\it Communications in Statistics - Simulation and Computation},  {\bf 17}, 581--607.
 
 
 
  \bibitem{}  Hotelling, H. (1936). Relations between two sets of variables. {\it Biometrika}, {\bf 58},  433--451. 
 
 \bibitem{}
Hsing, T. (1999). Nearest neighbor inverse regression. {\it The Annals of Statistics}, {\bf 27}, 697--731. 
 
 
 \bibitem{}
 Izenman, A. J. (1975). Reduced-rank regression for the multivariate linear
model. {\it Journal of Multivariate Analysis}, {\bf 5}, 248--264.  
 
\bibitem{}
Li, B. (2018).
 {\it Sufficient Dimension Reduction: Methods and Applications with R}.
CRC Press.

\bibitem{}
Li, B. and Wang, S. (2007). On directional regression for dimension reduction. \textit{Journal of American Statistical Association},  {\bf 479}, 997--1008.

\bibitem{}
Li, B., Wen, S. and Zhu, L. X. (2008). On a projective resampling method
for dimension reduction with multivariate responses. \textit{Journal of American Statistical Association}, {\bf 103}, 1177--1186.

\bibitem{}
Li, B., Zha, H., and Chiaromonte, F. (2005). Contour regression: a general approach to dimension
reduction. {\it The Annals of Statistics}, {\bf 33}, 1580--1616.

\bibitem{} Li, K. C. (1991). Sliced inverse regression for dimension
reduction (with discussion). {\it Journal of the American
Statistical Association}, {\bf 86}, 316--342.



\bibitem{} Li, K. C. (1992). On principal Hessian directions for data visualization and dimension reduction: another application of Stein's Lemma. {\it Journal of the American
Statistical Association}, {\bf 87},
1025--1039.

\bibitem{} Li, K. C. and Duan, N. (1989). Regression analysis under
link violation. {\it The Annals of Statistics}, {\bf 17}, 1009--1052. 
 
  \bibitem{}
Lue H. H. (2009). Sliced inverse regression for multivariate response regression. {\it Journal of Statistical Planning and Inference}, {\bf 139}, 2656--2664. 
 
 \bibitem{}
Lue H. H. (2010). On principal Hessian directions for multivariate
response regressions. {\it  Computational  Statistics}, {\bf 25}, 619--632. 

 \bibitem{}
Luo, W. (2022). On efficient dimension reduction with respect to the interaction between two response variables.  {\it Journal of the Royal Statistical 
Society, Series B}. To appear. $https://doi.org/10.1111/rssb.12477$.
 
\bibitem{}
Ma, Y. and Zhu, L. P. (2013). A review on dimension reduction. {\it International Statistics Review}, {\bf 81}, 134--150. 
 
 \bibitem{}
Naik, P. and Tsai, C. L. (2000). Partial least squares estimator for single-index models. {\it Journal of the Royal Statistical Society, Series B},  {\bf 62},
763--771. 
 
\bibitem{}
Saracco, J. (2005). Asymptotics for pooled marginal slicing estimator based on $\rm{SIR}_\alpha$ approach. {\it Journal of Multivariate Analysis}, {\bf  96}, 117--135. 

 \bibitem{}
Setodji, C. M.  and Cook, R. D. (2004). K-means inverse regression.
{\it Technometrics}, {\bf 46}, 421--429.


\bibitem{} Sheng, W. and Yin, X. (2013). Direction estimation in single-index models
via distance covariance. \textit{Journal of Multivariate Analysis}, {\bf 122}, 148--161.

\bibitem{} Sheng, W. and Yin, X. (2016). Sufficient dimension reduction via distance covariance.
\textit{Journal of Computational and Graphical Statistics}, {\bf 25}, 91--104.

\bibitem{}
Sz{\'e}kely, G. J., Rizzo, M. L. and  Bakirov, N. K.  (2007). Measuring and testing dependence by correlation of distances. {\it The Annals of Statistics}, {\bf 35}, 2769--2794.
 
\bibitem{} 
Wang, P., Yin, X., Yuan, Q. and Kryscio, R. (2021). Feature filter for estimating central mean subspace and its sparse solution. {\it Computational Statistics and Data Analysis}, {\bf 163}, 107285. 
 
\bibitem{}
Wen, X. M., Setodji, C. M. and Adekpedjou, A. (2009). A minimum discrepancy approach to multivariate
dimension reduction via k-means inverse
regression. {\it Statistics and Its Interface}, {\bf 2}, 503--511.    
 
 
 \bibitem{} 
 Weng, J. and Yin, X. (2018). Fourier transform approach for inverse dimension
reduction method. {\it Journal of Nonparametric Statistics}, {\bf 30}, 1049--1071.  
 
  \bibitem{}  
Xia, Y., Tong, H., Li, W. K. and Zhu, L. X. (2002). An adaptive estimation of dimension reduction space (with discussion).  {\it Journal of the Royal Statistical 
Society, Series B}, {\bf 64}, 363--410.  
 
 \bibitem{}  
Xue, Y., Wang, Q. and Yin, X. (2018). A unified approach to sufficient dimension reduction. {\it Journal of Statistical Planning and Inference}, {\bf 197}, 168--179. 
 
 \bibitem{} 
Yin, X. (2010). Sufficient dimension reduction in regression. In
{\it High-Dimensional Data Analysis}  Eds. X. Shen and T. Cai., 257--273. Singapore:
World Scientific.

 \bibitem{} 
Yin, X., Bura, E. (2006).  Moment-based dimension reduction for multivariate response regression. {\it Journal of Statistical Planning and Inference}, {\bf 136}, 3675--3688.

\bibitem{} Yin, X. and  Cook,   R. D. (2002). Dimension reduction for the
conditional $k$-th moment in regression. {\it Journal of the Royal Statistical Society, Series B},  {\bf 64}, 159--175.


\bibitem{} Yin, X. and Li, B. (2011). Sufficient dimension reduction based on an ensemble of minimum average variance estimators. {\it The Annals of Statistics}, {\bf 39}, 3392--3416.

\bibitem{} Ying, C. and Yu, Z. (2020). Fr{\'e}chet  sufficient dimension reduction for random objects. Submitted. $https://arxiv.org/abs/2007.00292$.

\bibitem{} Yoo, J. K. and Cook, R. D. (2007). Optimal sufficient dimension reduction for the conditional mean in multivariate
regression. {\it Biometrika}, {\bf 94}, 231--242.

\bibitem{} Yoo, J. K., Lee, K. and Wu, S. (2010).
On the extension of sliced average variance estimation
to multivariate regression. {\it Statistical Methods and Applications}, {\bf 19}, 529--540. 

\bibitem{}
Zhang, H. F. (2021). Minimum average variance estimation with group lasso for the multivariate response central mean subspace. {\it Journal of Multivariate Analysis}, {\bf 184}, 104753.

\bibitem{}
Zhang, Y., Zhu, L. P. and Ma, Y. (2017). Efficient dimension reduction for multivariate response data. {\it Journal of Multivariate Analysis}, {\bf 155}, 187--199.

\bibitem{} Zhou, J. and He, X. (2008). Dimension reduction based on constrained canonical correlation and variable filtering. {\it The Annals of Statistics}, {\bf  36}, 1649--1668. 

\bibitem{} 
Zhu, L. P. and Zhong, W. (2015). Estimation and inference on central mean subspace for multivariate response data. {\it Computational Statistics and Data Analysis}, {\bf 92}, 68--83.

\bibitem{} 
Zhu, L. P., Zhu, L. X. and Wen, S.  (2010). On dimension reduction in regressions with multivariate responses. {\it Statistica Sinaca}, {\bf 20}, 1291--1307.


\bibitem{} 
Zhu, Y. and Zeng, P. (2006). Fourier methods for estimating the central subspace and the central
mean subspace in regression. {\it Journal of the American Statistical Association},  {\bf 101}, 1638--1651.

\end{thebibliography}
 \end{document}